\documentclass{ws-mpla}
\usepackage[super]{cite}
\usepackage{graphicx,color}
\allowdisplaybreaks

\newcommand{\red}{\textcolor{red}}

\begin{document}


\markboth{Dev, Mohapatra, Zhang}
{Heavy right-handed neutrino dark matter in left-right models}

\catchline{}{}{}{}{}

\title{Heavy right-handed neutrino dark matter in left-right models}

\author{P. S. Bhupal Dev}

\address{Max-Planck-Institut f\"{u}r Kernphysik, Saupfercheckweg 1, D-69117 Heidelberg, Germany and \\
Department of Physics and McDonnell Center for the Space Sciences, \\ Washington University, St. Louis, MO 63130, USA}

\author{Rabindra N. Mohapatra}

\address{Maryland Center for Fundamental Physics, \\ Department of Physics, University of Maryland, College Park, MD 20742, USA}

\author{Yongchao Zhang}

\address{Service de Physique Th\'{e}orique, Universit\'{e} Libre de Bruxelles, \\ Boulevard du Triomphe, CP225, 1050 Brussels, Belgium}

\maketitle


\begin{abstract}
We show that in a class of non-supersymmetric left-right extensions of the Standard Model (SM), the lightest right-handed neutrino (RHN) can play the role of thermal Dark Matter (DM) in the Universe for a wide mass range from TeV to PeV. Our model is based on the gauge group $SU(3)_c \times SU(2)_L\times SU(2)_R\times U(1)_{Y_L}\times U(1)_{Y_R}$ in which a heavy copy of the SM fermions are introduced and the stability of the RHN DM is guaranteed by an automatic $Z_2$ symmetry present in the leptonic sector. In such models the active neutrino masses are obtained via the type-II seesaw mechanism. We find a lower bound on the RHN DM mass of order TeV from relic density constraints, as well as an unitarity upper bound in the multi-TeV to PeV scale, depending on the entropy dilution factor.
The RHN DM could be made long-lived by soft-breaking of the $Z_2$ symmetry and provides a concrete example of decaying DM interpretation of the PeV neutrinos observed at IceCube.

\end{abstract}


\section{Introduction}
The existence of dark matter (DM) in our Universe has been well established by astrophysical and cosmological observations. To explain this, one needs some new particle(s) and symmetries beyond the Standard Model (SM). Being colorless and electrically neutral, SM-singlet right-handed neutrino (RHN), introduced in seesaw models to explain the small neutrino mass, is a natural DM candidate. However in the conventional seesaw models\cite{type1a, type1b, type1c, type1d, type1e}, its Yukawa couplings to SM lepton and Higgs doublets make it highly unstable, and therefore, unsuitable for DM, unless its mass is in the keV range with appropriately small Yukawa couplings\cite{Adhikari:2016bei}. In ultraviolet (UV) complete scenarios of seesaw, e.g. those based on the Left-Right (LR) gauge group $SU(2)_L \times SU(2)_R \times U(1)_{B-L}$~\cite{LR1, LR2, LR3}, the gauge interactions of RHNs induce further decay modes, which have to be suppressed/forbidden by imposing additional discrete symmetries (or making the model more contrived) to keep the lightest RHN cosmologically stable\cite{Fiorentin:2016avj}.

In this paper we discuss an alternative class of LR models based on the gauge group $SU(3)_c \times SU(2)_L\times SU(2)_R\times U(1)_{Y_L}\times U(1)_{Y_R}$~\cite{Dev:2016xcp} which has an accidental $Z_2$ symmetry, at the renormalizable level that keeps the lightest RHN stable making it a natural DM candidate. In this model there exists a right-handed copy of the SM electroweak sector at the TeV scale (or higher), charged under $SU(2)_R \times U(1)_{Y_R}$. The breaking of this gauge symmetry down to $U(1)_{\rm em}$ is similar to the LR models with universal seesaw\cite{Babu:1989rb,CP1,CP2}. But one of the key differences from the universal seesaw models is that the right-handed lepton sector in our model is prevented by gauge symmetry from having any masses connecting the heavy sector to the light ones. Because of this, there appears a remnant automatic $Z_2$ symmetry in the leptonic sector at the renormalizable level that keeps the lightest RHN stable.
If the stabilizing $Z_2$ symmetry is softly broken by a mass term connecting the heavy and light sector, the RHN DM can be made very long-lived but allowing for its decay into SM leptons. 
This raises the possibility that for a PeV-scale DM, one can explain the ultra-high energy neutrino events observed at IceCube~\cite{Dev:2016qbd}. We also point out a dilution mechanism which allows us to relax the unitarity bound on the dark matter to the PeV mass level so that it can be applied to the discussion of PeV neutrinos in the context of a thermal annihilating DM.

\section{The Model}
\label{sec:model}

The model is based on the gauge group $SU(2)_L \times SU(2)_R \times U(1)_{Y_L} \times U(1)_{Y_R}$.~\cite{BCS,BCS2} For each generation of SM fermions
\begin{eqnarray}
&& Q_{L} \ \equiv \ \left(\begin{array}{c} u\\ d\end{array}\right)_{L}: \bigg({\bf 2}, {\bf 1}, \frac{1}{3}, 0\bigg), \quad
\psi_{L} \ \equiv \ \left(\begin{array}{c} \nu \\ e \end{array}\right)_{L}: ({\bf 2}, {\bf 1}, -1,0) , \nonumber \\
&& u_R: \bigg({\bf 1}, {\bf 1}, \frac{4}{3}, 0\bigg),~~
d_R: \bigg({\bf 1}, {\bf 1}, -\frac{2}{3}, 0\bigg),~~
e_R: ({\bf 1}, {\bf 1}, -2, 0) ,
\end{eqnarray}
we have the heavy fermion partners
\begin{eqnarray}
&& \mathcal{Q}_{R} \ \equiv \ \left(\begin{array}{c} \mathcal{U} \\ \mathcal{D} \end{array}\right)_{R}: \bigg({\bf 1}, {\bf 2}, 0, \frac{1}{3}\bigg), \quad
\Psi_{R} \ \equiv \ \left(\begin{array}{c} N \\ \mathcal{E} \end{array}\right)_{R}: ({\bf 1}, {\bf 2}, 0, -1), \nonumber \\
&& \mathcal{U}_L:\bigg({\bf 1}, {\bf 1}, 0, \frac{4}{3}\bigg),~~ \mathcal{D}_L:\bigg({\bf 1}, {\bf 1}, 0, -\frac{2}{3}\bigg),~~
\mathcal{E}_L:({\bf 1}, {\bf 1}, 0,-2) \, .
\end{eqnarray}
The Higgs sector of the model consists of $SU(2)_{L,R}$ doublets and triplets:
\begin{align}
& \chi_L \ = \ \left( \begin{matrix} \chi_L^{+} \\ \chi_L^0 \end{matrix} \right)  : ({\bf 2}, {\bf 1}, 1, 0) \,, \qquad  
\chi_R \ = \ \left( \begin{matrix} \chi_R^{ \, +} \\ \chi_R^{\, 0} \end{matrix} \right)  : ({\bf 1}, {\bf 2}, 0, 1) \,, \nonumber \\
& \Delta_L \ = \ \left(\begin{matrix}\frac{\Delta^+_L}{\sqrt{2}} & \Delta^{++}_L\\ \Delta^0_L & -\frac{\Delta^+_L}{\sqrt{2}}\end{matrix}\right)  : ({\bf 3}, {\bf 1}, 2, 0) \,,  \qquad 
\Delta_R \ = \ \left(\begin{matrix}\frac{\Delta^+_R}{\sqrt{2}} & \Delta^{++}_R\\ \Delta^0_R & -\frac{\Delta^+_R}{\sqrt{2}}\end{matrix}\right)  : ({\bf 1}, {\bf 3}, 0, 2) \,.
\end{align}
The SM $SU(2)_L$ and new $SU(2)_R$ gauge symmetries are broken by the vacuum expectation values (VEV) of the neutral components of these Higgs fields:
\begin{align}
\langle \chi_L^0 \rangle  =  v_{L} \equiv v_{\rm EW}, \qquad \langle \chi_R^{0} \rangle  =  v_R, \qquad \langle \Delta_L^0 \rangle  =  w_L,\qquad  \langle \Delta_R^0 \rangle  =  w_R.
\end{align}
 The SM and heavy charged fermion masses are obtained through their Yukawa couplings to the Higgs doublets:
\begin{eqnarray}
\label{eq:Lyukawa}
- \mathcal{L}_Y & \ \supset \ &
 y_{u}  \overline{Q}_L \widetilde{\chi}_{L} u_R
+ y_{d}\overline{Q}_L  \chi_{L} d_R
+ y_{e} \overline{\psi}_L  \chi_{L} e_R \nonumber \\
&& +y'_{u}  \overline{\cal Q}_R \widetilde{\chi}_{R} {\cal U}_L
+ y'_{d}\overline{\cal Q}_R  \chi_{R} {\cal D}_L
+ y'_{e} \overline{\Psi}_R  \chi_{R} {\cal E}_L
+ {\rm H.c.} \,,
\end{eqnarray}
where $\tilde{\chi}_L = i\sigma_2\chi_L^*$  and similarly for $\chi_R$ ($\sigma_2$ being the second Pauli matrix). 
Exact parity symmetry implies that the Yukawa couplings $y_f = y'_f$ (and also $f = f'$ in Eq.~(\ref{eq:Lyukawa2}) below). If the $SU(2)_R$ symmetry breaking is at the TeV scale, there would be new quarks in the few GeV range, which is inconsistent with the current LHC bounds on vector-like fermion masses of order of TeV. To have heavy quarks and leptons at the TeV scale, we adopt the parity violating scenario in which the couplings $y_f$ and $y'_f$ are independent and choose $y'_f \sim {\cal O} (1)$.\footnote{The parity symmetric version could be realized at the TeV scale by doubling the doublet scalars in both the SM and heavy sectors, such that one set of the Yukawa couplings $y_{f,1}$ is responsible for the SM fermion masses and the other one $y_{f,2}$ dominates the heavy fermion masses~\cite{Dev:2016xcp}.} It is worth noting that at this stage the lower bound on $M_{W_R}$ from low-energy flavor changing effects such as $K^0-\overline{K^0}$ mixing does not apply to our model.

The masses of light and heavy neutrinos are generated from couplings to the triplet scalars $\Delta_{L,R}$ via the type-II seesaw mechanism~\cite{type2a, type2b, type2c, type2d}:
\begin{eqnarray}
\label{eq:Lyukawa2}
- \mathcal{L}_Y & \ \supset \ &
f \overline{ \psi_L^C} i\sigma_2 \Delta_L \psi_L 
+ f' \overline{ \Psi_R^C} i\sigma_2 \Delta_R \Psi_R
+ {\rm H.c.} \,,
\end{eqnarray}
which implies that the VEV $w_L$ should be at the eV scale and $w_R$ is at the TeV scale or above. This can be easily obtained by choosing appropriate values of the parameters in the scalar potential.

\section{Depleting the heavy quarks}

After symmetry breaking at the right-handed scale and the electroweak scale, the model has a large global symmetry: $U(1)_{B,\,L}\times U(1)_{B,\,R}\times Z_{2\ell,\,L} \times Z_{2 \ell,\, R}$, with the two $U(1)_B$ the baryon numbers in the SM and heavy sectors, and the two $Z_2$'s the corresponding lepton numbers. Protected by the baryon number symmetry, the lightest hadron in the $SU(2)_R$ sector is absolutely stable, which can be one of the heavy baryons of form $QQQ$, $QQq$, $Qqq$, or heavy mesons $\bar{Q}Q$ and $\bar{Q}q$.

For the phenomenological purpose of avoiding the heavy lightest baryon also becoming DM and affecting the evolution of the Universe, we add two more Higgs singlets $\Sigma_1({\bf 1},{\bf 1}, -\frac43, +\frac43)$ and $\Sigma_2({\bf 1},{\bf 1}, +\frac23,- \frac23)$ which connect the heavy and light singlet quarks via the terms
\begin{eqnarray}
\label{eqn:Lmix}
{\cal L}_{\rm mix} \ = \ \lambda_U \overline{\mathcal{U}}_Lu_R\Sigma^*_1+\lambda_D\overline{\mathcal{D}}_Ld_R\Sigma^*_2+ {\rm H.c.} \,.
\end{eqnarray}
Once the $\Sigma$ fields acquire VEVs, the heavy-light quark mass mixing terms $\delta_{U} = \lambda_{U} \langle \Sigma_1 \rangle$ and $\delta_D = \lambda_D \langle \Sigma_2 \rangle$ appear in the Lagrangian, and the heavy quark can then decay into the SM fermions, leaving only the lightest RHN stable due to the leptonic $Z_2$ symmetry. The singlets $\Sigma_{1,2}$ are charged under $U(1)_{Y_L} \times U(1)_{Y_R}$ and break this symmetry down to $U(1)_{\cal Y}$. At this stage this model is similar to the usual LR models with $U(1)_{B-L}$, however with the quantum number $\cal Y$ different from $B-L$ and also from the SM hypercharge $Y_L$ (which is a linear combination of $\cal Y$ and the
$U(1)$ subgroup of $SU(2)_R$ after symmetry breaking).

The decay rate of heavy quarks $Q \to q + Z$ can be estimated as
\begin{eqnarray}
\Gamma_{Q} \ \simeq \ \frac{g_L^2}{64\pi \cos^2\theta_w} \frac{\delta^2_Q}{v_R^2} \frac{M_Q^3}{M_Z^2}\,,
\label{MQq}
\end{eqnarray}
where $\theta_w$ is the SM weak mixing angle and $g_L$ is the gauge coupling of $SU(2)_L$.
Requiring that the heavy quarks are depleted before the QCD phase transition epoch $T_{\rm QCD}\approx 200$ MeV imposes a lower bound on the magnitude of $\delta_{Q} \gtrsim 10^{-6}$ GeV~\cite{Dev:2016xcp}. When this limit is close to being saturated, we expect the heavy quark decay length to be several cm, leading to displaced vertex signatures at hadron colliders.

\section{Connecting the SM to DM}
\label{sec:2.4}

In absence of any scalar and fermion mixings connecting the heavy and light sectors, the RHN $N$ interacts \red{pairwise} with the SM fermions only through the heavy $Z_R$ boson, which couples directly to the SM sector through the $U(1)_{\mathcal Y}$ interaction. Once the mixing terms are included, at the lowest order, two $N$'s could also annihilate to the SM fields via the SM Higgs $h$ and $\Delta_R^0$ as well as the $Z$ boson, induced due to the $h - \Delta_R^0$ and $Z - Z_R$ mixings at tree level. The SM Higgs $h$ is assumed to be predominantly from the doublet $\chi_{L}$, and thus, the scalar mixing is expected to be of order $\zeta_S \simeq \lambda v_{\rm EW} / w_R$ from the quartic term $\lambda ( \chi_L^\dagger \chi_L ) {\rm Tr} ( \Delta_R^{\dagger} \Delta_R )$, where for simplicity we have neglected the mixing with other scalars. The neutral gauge boson mixing is of order $\zeta_Z\simeq \xi_Z(M_Z/M_{Z_R})^2$, where $\zeta_Z=\sin\theta_w(\cot^2\theta_w g_R^2/g_L^2-1)^{-1/2}$.
Note that when the gauge coupling $g_R$ approaches the theoretical lower limit~\cite{rizzo} $g_L \tan\theta_w$ (which is independent of the symmetry breaking pattern~\cite{Dev:2016dja}), the $Z - Z_R$ mixing could be significantly enhanced.

In the minimal version of our model, we do not have the bi-multiplet scalars which transform non-trivially under both $SU(2)_L$ and $SU(2)_R$, thus the charged gauge bosons $W$ and $W_R$ can not mix at the tree level. A tiny mixing arises at one-loop level, induce by the heavy-light quark mixings, which is estimated to be $\zeta_{W} \sim 10^{-12}$ for the third generation quarks, and even smaller for the first and second generations.

\section{DM annihilation and relic density}
\label{sec:3.2}

In the early universe, all the particles are in equilibrium due to the common $SU(3)_c$ color and $U(1)_{\mathcal Y}$ interactions.  As the universe cools down below the DM mass $M_N$, the DM density goes down and freezes out, as in case of a generic thermal DM candidate. The primary annihilation channels to the SM particles proceed via the scalar and the neutral gauge boson portals:
\begin{eqnarray}
NN \ \to \ \left\{\begin{array}{c} h^{(\ast)} / \Delta_R^{0 \, (\ast)} \\ Z^{(\ast)} / Z_R^{(\ast)}  \end{array} \right . \ \to \  \text{SM particles} \,.
\label{eq:ann}
\end{eqnarray}
The $h$ and $Z$ portals are suppressed respectively by the small mixing angles $\zeta_S$ and $\zeta_Z$. On the other hand, though the couplings of $Z_R$ to $N$ and the SM fermions are of order one, the $Z_R$ portal is suppressed by the large $Z_R$ mass, except for the Breit-Wigner resonance when $2M_N\simeq M_{Z_R}$. The $\Delta_R$ portal is suppressed when DM is light, as the Yukawa coupling $f'$ is proportional to the DM mass via $M_N = 2 f' w_R$. Combining all the available channels, we calculate the thermally averaged DM annihilation cross section times velocity $\langle \sigma v \rangle = a + b \langle v^2 \rangle + \mathcal{O} (v^4)$, from which the current DM relic density can be obtained via
\begin{eqnarray}
\Omega_N h^2 \ = \ \frac{1.07 \times 10^9 \, {\rm GeV}^{-1}}{M_{\rm Pl}} \frac{x_F}{\sqrt{g_\ast}} \frac{1}{a+3b/x_F} \,,
\label{eq:relic1}
\end{eqnarray}
where $x_F = M_N / T_F \simeq 20$ (with $T_F$ being the freeze-out temperature) and $g_\ast = 106.75$ is the relativistic degrees of freedom at $T_F$.

\begin{figure}[t!]
  \centering
  \includegraphics[width=0.6\textwidth]{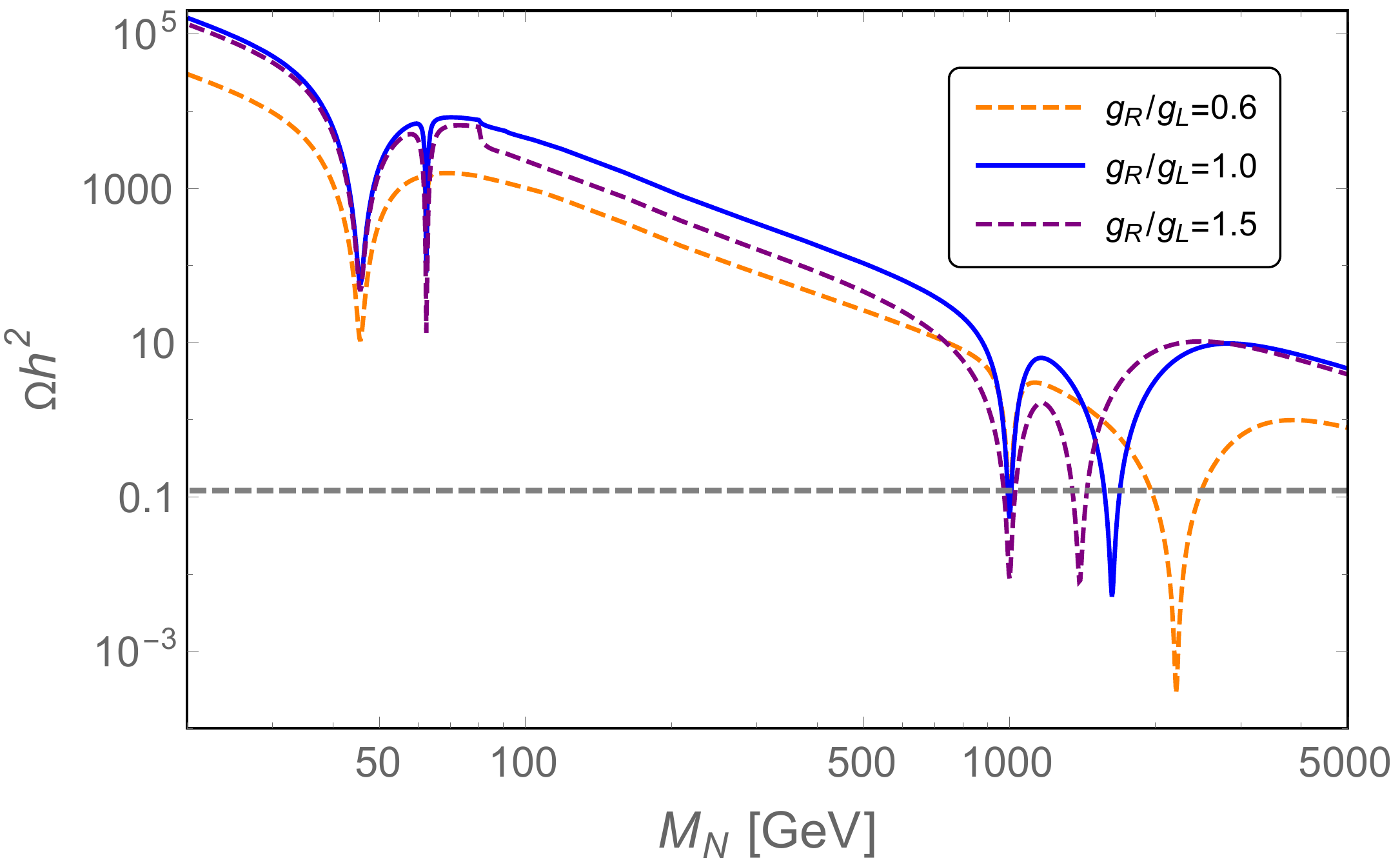} \\
  \caption{An illustration of the relic density of RHN DM as a function of its mass $M_N$ for different values of $g_R/g_L$. The horizontal line gives the observed value from Planck data~\cite{Ade:2015xua}. }
  \label{fig:density}
\end{figure}



An example is given in Fig.~\ref{fig:density}, where we set explicitly the quartic coupling $\lambda = 1$ and the heavy scalar mass $M_{\Delta_R^0} = 2$ TeV with width 30 GeV. Three different value of the gauge coupling $g_R$ are chosen with $g_R / g_L = 0.6$, 1 and 1.5. The corresponding $Z_R$ mass and the RH scale $v_R = w_R$ are set to their current experimental lower bounds~\cite{Aaboud:2016cth}, viz. $M_{Z_R}=4.4, \, 3.3, \, 2.8$ TeV and $w_R=2.9, \, 2.6, \, 1.7$ TeV for $g_R/g_L=0.6, \, 1, \, 1.5$  respectively.
The horizontal dashed line shows the observed relic density, as measured by Planck~\cite{Ade:2015xua}. The various peaks in Fig.~\ref{fig:density} are respectively (from left to right) due to the SM $Z$ and Higgs bosons, $\Delta_R^0$ and $Z_R$.
We find that a TeV scale RHN could accommodate the observed DM relic density, with the annihilation dominated by the heavy scalar and/or $Z_R$ bosons, depending largely on the quartic coupling $\lambda$, the gauge coupling $g_R$ and the VEV $v_R$.


\section{Going beyond the unitarity bound}

\begin{figure}[t!]
  \centering
  \includegraphics[width=0.6\textwidth]{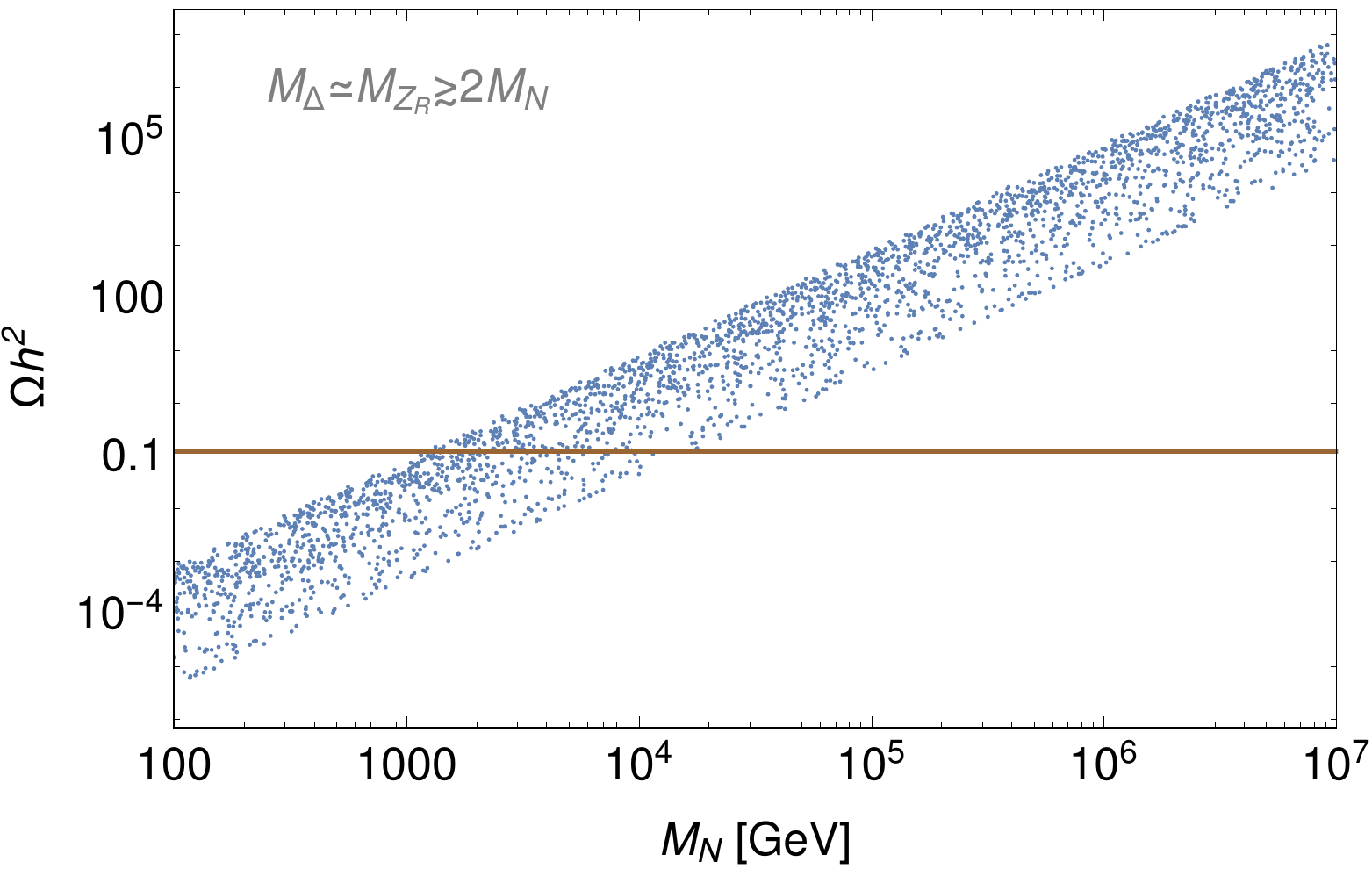} \\
  \caption{The relic density of RHN DM as a function of its mass $M_N$ at the resonance for different values of $\lambda$. The horizontal line gives the observed value from Planck data~\cite{Ade:2015xua}. }
  \label{fig:uni}
\end{figure}

In generic annihilating thermal DM models, the DM mass cannot be arbitrarily large due to the well-known partial wave unitarity limit~\cite{kamion}. To see this explicitly in our model, we write down the thermal averaged annihilation cross section at leading order in $v_{\rm EW}^2 / v_R^2$:
\begin{align}
\label{eqn:sigmav_pev}
\langle \sigma v \rangle
\ = \ & \frac{3 f^{\prime 2} \lambda^2  \langle v^2 \rangle}{64 \pi v_R^2} \frac{M_N^4}{(4M_N^2-M_\Delta^2)^2+M_\Delta^2\Gamma_\Delta^2}
+ \frac{\tilde{g}^4 M_N^2}{64 \pi M_{Z_R}^4}
\left( 1 - \frac14 \langle v^2 \rangle \right) \nonumber \\
& + \frac{(5 g_R^4 \tan^4\phi+\tilde{g}^4) \langle v^2 \rangle}{24\pi}
\frac{M_N^2}{(4M_N^2-M_{Z_R}^2)^2+M_{Z_R}^2\Gamma_{Z_R}^2} \,,
\end{align}
where $\tilde g^4 \equiv g_L^2 g_R^2 \xi_Z^2 / (16 \cos^2 \theta_w \cos^2 \phi)$. 
Thus the cross section is suppressed by the right-handed scale $v_R$, i.e. $\langle \sigma v \rangle \propto v_R^{-2}$, except when $2M_N\simeq M_{\Delta}$ or $M_{Z_R}$, which results in a Breit-Wigner enhancement. Even in this case, we must ensure that the maximum value of the cross section at the resonance obeys the partial wave unitarity limit~\cite{Nussinov:2014qva}. For the resonance $R(=\Delta, Z_R)$ just above the threshold, Eq.~\eqref{eqn:sigmav_pev}  can be approximated by
\begin{eqnarray}
\langle \sigma v \rangle \ \simeq \ 16 \pi \, \frac{\Gamma_{NN}\Gamma_{\rm SM}}{(4M^2_N-M^2_R)^2+\Gamma^2_{R}M^2_R}
\ \sim \  \frac{4\pi} {M_N^2} \left( B_{NN}B_{\rm SM} \right) \,,
\label{eq:uni2}
\end{eqnarray}
where $\Gamma_{NN}$ and $\Gamma_{\rm SM}$ are the partial decay widths of $R$ to the DM pair and SM particles respectively, and $B_{NN}$ and $B_{\rm SM}$ are the corresponding branching ratios. From Eq.~\eqref{eq:uni2}, we find that the annihilation rate decreases with increasing DM mass, which leads to the unitarity bound. This is shown in Fig.~\ref{fig:uni} where we have fixed the $\Delta$ and $Z_R$ masses at the resonance and have varied the coupling $\lambda$ between $10^{-5}$ and $\sqrt{4\pi}$ to obtain the scattered points for the relic density as a function of the DM mass. We find that only the $M_N$ values between 1--20 TeV can explain the observed relic density in this scenario.

It turns out that there is a built-in mechanism for entropy dilution in our model, thus making it possible to go beyond the unitarity limit.  The basic idea is that the next-to-lightest heavy neutrino (denoted here by $N_2$) does not need to be stable and can decay into SM leptons, e.g. $N_2 \to \ell \ell \nu$ via the charged lepton mixing terms in Eq.~(\ref{eqn:Lmix2}) involving only the heavier than $N$ leptons in the $SU(2)_R$ sector. Note that the hadronic channels $N_2 \to \ell q \bar q$ are comparatively further suppressed by the small quark mixing $\delta_Q$, and the two-body decay $N_2 \to W^\pm \ell^\mp$ is heavily suppressed by the loop-induced $W - W_R$ mixing.
Because of higher mass, $N_2$ freezes out before the DM $N$ but could decay after $N$ freezes out, i.e. $T_{N_2}< T_F$, with appropriate values of leptonic mixing.
The decay of $N_2$ to relativistic SM species will generate entropy which dilutes the relic density of $N$, and therefore, the initial relic density of $N$ can be significantly larger than the current value, thus allowing for larger $M_N$ values in Fig.~\eqref{fig:uni}. 

The $N_2$ decay temperature is given by
\begin{eqnarray}
T_{N_2} \ \simeq \ 0. 78 g_\ast^{-1/4}\sqrt{\Gamma_{N_2}M_{\rm Pl}} \,,
\end{eqnarray}
where $M_{\rm Pl}$ is the Planck scale and the decay width
\begin{eqnarray}
\Gamma ( N_{2} \to \ell \ell \nu ) \ \simeq \
\frac{y_e^2 y_e^{\prime 2} N_{\rm lepton}}{96 \pi^3} \frac{M_{N_2}^5}{M_\Delta^4}
\left( \frac{\delta_\ell}{v_R} \right)^2 \delta_{S}^2 \,, 
\end{eqnarray}
with $N_{\rm lepton}$ the number of degrees of leptons, $\delta_\ell$ the heavy-light charged lepton mixing in Eq.~(\ref{eqn:Lmix2}), and $\delta_{S}$ the charged scalar mixing~\cite{Dev:2016qbd} (without the assumption of lepton-specific structure of the scalar sector).
By adjusting the mixing parameters $\delta_{S}$ and $\delta_\ell$, we can make $N_2$ decay after $T_F$.

Now we calculate the relic density of $N_2$ (mostly governed by the off-resonance annihilation $N_2N_2\to $ SM fields), which will determine the extent of dilution. We find the number density of $N_2$ (normalized to the entropy density $s$)
\begin{eqnarray}
Y_{N_2} \ \simeq \ \frac{M_{N_2}}{\delta_{N_2} T_{N_2}}
\end{eqnarray}
where $\delta_{N_2}=\frac{2\pi^2g_*}{45}\frac{M^3_{N_2} \langle \sigma v \rangle }{H(M_{N_2})}$. Using for instance the scalar channel with $\langle \sigma v \rangle \simeq \frac{3\lambda^2f^{\prime 2}}{1024 \pi M^2_{N_2}} \langle v^2 \rangle$, we find $Y_{N_2}\simeq 10^{-2}$ for $M_{N_2}\sim 10$ PeV.

The dilution factor is then calculated by equating the energy density before and after the decay:
\begin{eqnarray}
Y_{N_2}M_{N_2} s_{\rm before} \ = \ \frac{3}{4}s_{\rm after}T_{N_2} \,.
\end{eqnarray}
The dilution factor is then given by $d = {s_{\rm after}}/{s_{\rm before}}$, i.e.
\begin{eqnarray}
d \ = \ \frac{4}{3}~\frac{Y_{N_2}M_{N_2}}{T_{N_2}} \,.
\end{eqnarray}
With mixing parameters $\delta_S \simeq 0.1$, $\delta_\ell \simeq 0.01$ GeV, $\lambda \simeq 0.03$ and $y_e^{(\prime)} \simeq 0.3$, we can get a dilution factor as big as $d\sim 10^6$, which is adequate to give the right relic density of $N$ for $M_N$ up to a PeV or so, as can be seen from Fig.~\ref{fig:uni}.

\section{Direct and indirect detections and collider searches} \label{sec:4}

As a Majorana DM candidate, the RHN has both spin-independent (SI) and spin-dependent (SD) interactions with nuclei, which are mediated by the scalars and neutral gauge bosons, respectively.

\begin{figure}[t!]
  \centering
  \includegraphics[width=0.6\textwidth]{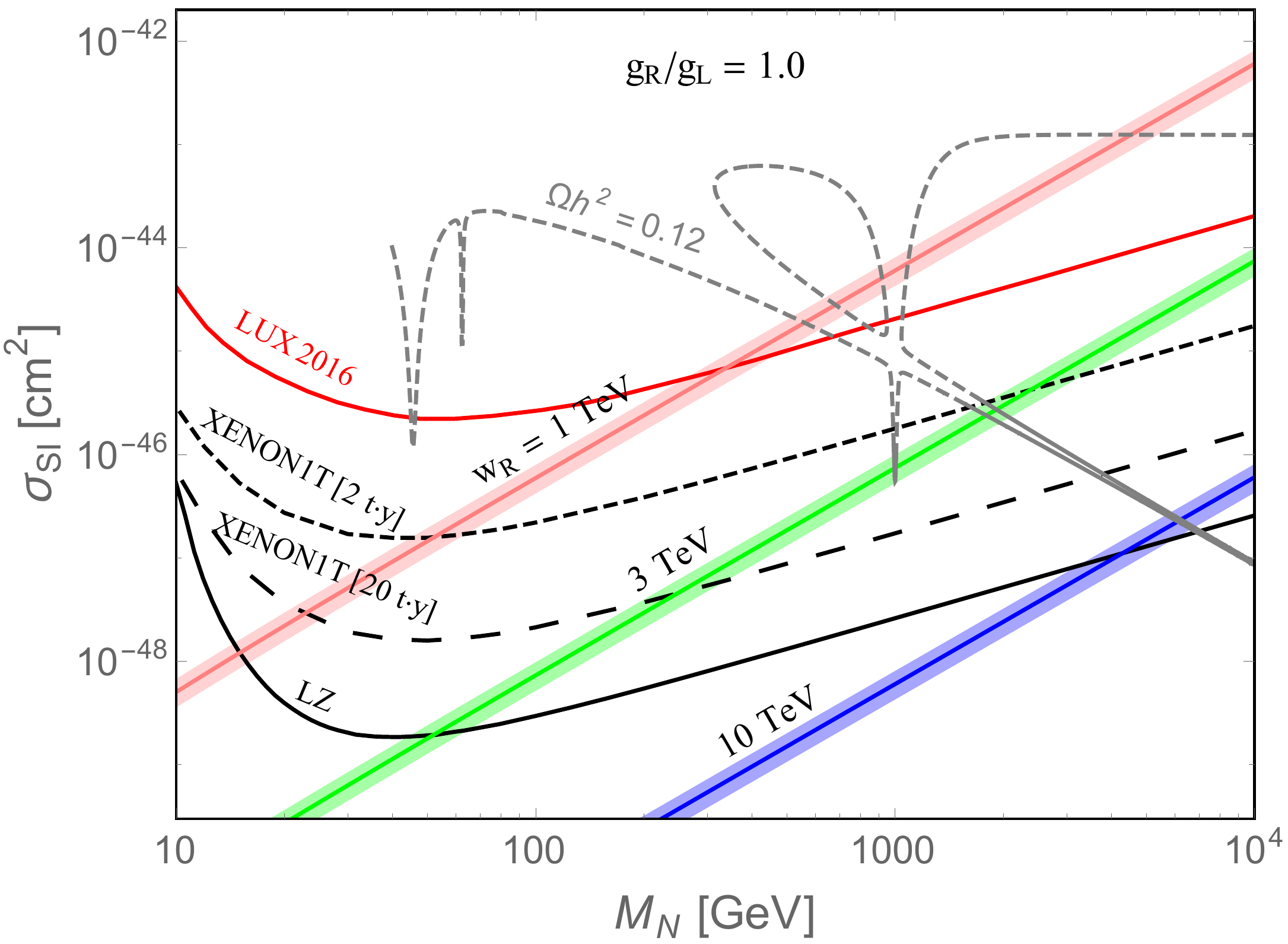}
  \caption{Predictions for the SI DM-nucleon scattering cross sections as functions of DM mass $M_N$ with the gauge coupling $g_R = g_L$ and the RH scale $w_R = 1$, 3 and 10 TeV. The colorful bands are due to the sizable uncertainties of effective DM-nucleon couplings $f_{p,\,n}$. We show also the current upper limits from LUX~\cite{Akerib:2016vxi}, as well as the future reaches of XENON1T~\cite{Aprile:2015uzo} and LZ~\cite{Akerib:2015cja}. The gray curves correspond to the parameter space producing the observed relic density $\Omega h^2 = 0.12$ (with the dashed part excluded by direct collider searches of $Z_R$ in the dilepton channel).}
  \label{fig:dd1}
\end{figure}

The amplitude due to heavy scalar exchange is suppressed by its large mass, thus the SI scattering cross section is dominated by the SM Higgs contribution
\begin{eqnarray}
\label{eqn:sigmaSI}
\sigma_{\rm SI} \ = \
\frac{\lambda^2 \mu^2 M_N^2}{2\pi m_h^4 w_{R}^4 }
\left[ Z f_p^{} + (A-Z) f_n^{} \right]^2
\end{eqnarray}
where $\mu$ is the reduced DM-nucleus mass, and $f_{p,n}$ the effective Higgs-proton/neutron couplings.
A  characteristic feature of our model is that $\sigma_{\rm SI} \propto M_N^2$, which originates from the dependence of Yukawa coupling on the DM mass $f' \propto M_N$. As shown in Fig.~\ref{fig:dd1}, when the right-handed scale $w_R$ goes higher, the scalar and gauge mixings are decreased and the scattering cross section $\sigma_{\rm SI}$ also gets smaller. In this plot we adopt the benchmark values of $v_R = w_R =$ 1, 3 and 10 TeV, the gauge coupling $g_R = g_L$ and the quartic coupling $\lambda = 1$.
The colorful bands are due to the sizable uncertainties of effective couplings $f_{p,\,n}$. The most stringent current direct detection limits come from LUX~\cite{Akerib:2016vxi} (with comparable limits from PandaX-II~\cite{Tan:2016zwf}), which rule out the lighter RHN DM mass range, if the DM relic density is set to be the observed value (the gray contours). When the collider limits on $Z_R$ mass are also taken into consideration (with the dashed gray lines excluded in Fig.~\ref{fig:dd1}), we can set more stringent lower limits of a TeV or so on the DM mass. We show also the projected limits from XENON1T with exposure values of 2 and 20 ton$\cdot$year~\cite{Aprile:2015uzo} and the LZ experiment~\cite{Akerib:2015cja} with orders of magnitude improvement in the SI scattering cross section limits, which can constrain the DM mass $M_N$ and right-handed scale $w_R$ up to the 10 TeV scale in near future experiments.


As for the SD scattering, it is mediated at leading order by the $Z$ and $Z_R$ bosons, both of which are suppressed by the small $Z - Z_R$ mixing angle $\zeta_Z$.
When the collider constraints on $M_{Z_R}$ are taken into consideration, the $Z - Z_R$ mixing is so small that even with the future LZ limits it is rather challenging to test the model in the SD channels.

In the early universe, the RHN DM annihilates into the SM particles, e.g. $W^+ W^-$, $\ell^+ \ell^-$ and $q \bar q$, which could emit  energetic gamma-rays. Thus the RHN DM can also be probed by the Fermi-LAT gamma-ray observations. Also DM annihilation injects energy into the thermal bath in the early universe, thus the annihilation cross section could also be constrained by the Planck data of the cosmic microwave background (CMB). However, limited by the small scalar and gauge mixing angles ($\zeta_S$ and $\zeta_Z$), these indirect detections do not put additional constraints on the currently allowed model parameter space with the observed relic density.

As in usual thermal DM models, the RHN DM can be pair-produced at high energy hadron colliders, with monojet, mono-$h$, mono-$W/Z$ and mono-photon signatures. However, suppressed again by the small mixing parameters, none of these channels could effectively probe our model at the LHC. It might be helpful to check some specific parameter regions at future higher energy colliders, such as FCC-hh and SPPC.


\section{Decaying RHN DM and IceCube PeV neutrinos}

In the model discussed in Section~\ref{sec:model}, the SM charged leptons $e_R$ do not mix with their heavy partners ${\cal E}_L$, thus the lightest RHN is stabilized by the automatic $Z_2$ symmetry. If we add the following term to the Lagrangian
\begin{eqnarray}
\label{eqn:Lmix2}
{\cal L}_{\rm mix} \ = \ \delta_\ell \bar{\mathcal{E}}_Le_R + {\rm H.c.} \,,
\end{eqnarray}
to explicitly break the $Z_2$ symmetry (albeit softly), then the lightest RHN can decay into SM lepton and $W$ boson through the loop mediated $W - W_R$ mixing, i.e. $N \to W^\pm \ell^\mp$. This raises the interesting possibility of explaining the IceCube PeV neutrinos~\cite{Aartsen:2015zva} when we set $M_N \sim$ few PeV~\cite{Dev:2016qbd}. However, the two-body decays are disfavored by the IceCube data, as the neutrino spectrum is flatter than required~\cite{Dev:2016qbd}.

To fit the IceCube neutrino data, we promote the scalar sector to that of a lepton-specific two Higgs doublet model (2HDM) in both the SM and heavy sectors, with the Yukawa couplings given by
\begin{eqnarray}
\label{eq:Lyukawa}
- \mathcal{L}_Y & \  \supset \ &
 y_{u}  \bar{Q}_L \tilde{\chi}_{L,q} u_R
+ y_{d}\bar{Q}_L  \chi_{L,q} d_R
+ y_{\ell} \bar{\psi}_L  \chi_{L,\ell} e_R
\nonumber \\
&& + y'_{u}  \bar{\cal Q}_R \tilde{\chi}_{R,q} {\cal U}_L
+ y'_{d}\bar{\cal Q}_R  \chi_{R,q} {\cal D}_L
+ y'_{\ell} \bar{\Psi}_R  \chi_{R,\ell} {\cal E}_L
+ {\rm H.c.} \,,
\end{eqnarray}
where the electroweak VEV $v_{\rm EW} = \sqrt{v_{L,\ell}^2 + v_{L,q}^2}$ with $v_{L,\ell} = \langle \chi_{L,\ell}^0 \rangle$ and $v_{L,q} = \langle \chi_{L,q}^0 \rangle$. When the VEV ratio $\tan\beta = v_{L,q} / v_{L,\ell} \gg 1$, then the singly-charged scalar is predominantly from the lepton doublet, i.e. $\chi_L^\pm \simeq (\chi_{L,\ell}^\pm - \cot\beta \chi_{L,q}^\pm)$, and its hadronic decays are highly suppressed by $\tan\beta$. Mediated by the mixing of $\chi_L^\pm$ with its heavy partner $\chi_R^\pm$ which couples to the DM, we can have three-body decay of DM, i.e. $N \to \ell^+ \ell^- \nu$, which can successfully fit the IceCube data~\cite{Dev:2016qbd}. 

\section{Conclusion}

In summary, we have discussed the phenomenology and cosmology of a new TeV-scale RHN DM candidate in the context of an $SU(3)_c\times SU(2)_L\times SU(2)_R\times U(1)_{Y_L}\times U(1)_{Y_R}$ model, where we introduced an $SU(2)$ analog of the electroweak interaction and matter contents but with a common QCD. After symmetry breaking, there is an automatic $Z_{2}$ symmetry in the leptonic sector that guarantees the stability of the lightest RHN, which plays the role of cold DM in the universe.
The DM relic density and direct detection constraints imply a lower limit of order 1 TeV for the RHN DM. We have also discussed a dilution mechanism which can push the RHN dark matter mass to the multi-PeV range (beyond the conventional unitarity limit) so that together with a weak breaking of the $Z_2$ symmetry, the RHN DM can in principle explain the PeV neutrinos observed at IceCube. 

\section*{Acknowledgment}
We thank Shmuel Nussinov for discussions and comments. This work of B.D. was partly supported by the DFG grant RO 2516/5-1. The work of R. N. M. is supported by National Science Foundation grant no. PHY1620074. Y. Z. would like to thank the IISN and Belgian Science Policy (IAP VII/37) for support.

\end{document}